\documentstyle[epsf]{elsart}

\begin{document}

\begin{frontmatter}

\title{Modeling electricity loads in California:\\ 
ARMA models with hyperbolic noise}

\author[IM]{J. Nowicka-Zagrajek\thanksref{KBN1}\thanksref{AUT}},
\author[HSC]{R. Weron\thanksref{KBN1}}
\address[IM]{Institute of Mathematics,
              Wroc{\l}aw University of Technology, 50-370 Wroc{\l}aw, Poland
}
\address[HSC]{Hugo Steinhaus Center for Stochastic Methods,
              Wroc{\l}aw University of Technology, 50-370 Wroc{\l}aw, Poland
}

\thanks[KBN1]{Research partially supported by KBN Grant PBZ-KBN 016/P03/01.}
\thanks[AUT]{Corresponding author. E-mail address: nowicka@im.pwr.wroc.pl}

\begin{abstract}
In this paper we address the issue of modeling electricity loads. After analyzing
properties of the deseasonalized loads from the California power market we fit an 
ARMA(1,6) model to the data. The obtained residuals seem to be independent but with 
tails heavier than Gaussian. It turns out that the hyperbolic distribution provides
an excellent fit.
\end{abstract}


\end{frontmatter}

\section{Introduction}

During the last decade we have witnessed radical changes in the structure of electricity 
markets world-wide. For many years it was argued convincingly that the electricity industry 
was a natural monopoly and that strong vertical integration was an obvious and
efficient model for the power sector. However, recently it has been recognized that 
competition in generation services and separation of it from transmission and distribution
would be the optimal long-term solution. Restructuring has been designed to foster 
competition and create incentives for efficient investment in generation assets
\cite{ICC98,masson99,wolfram99}. 

While the global restructuring process has achieved some significant successes, serious 
problems -- some predictable, others not -- have also arisen.
The difficulties that have appeared were partly due to the flaws in regulation and partly
to the complexity of the market. 

When dealing with the power market we have to bear in mind that electricity cannot simply 
be manufactured, transported and delivered at the press of a button. Moreover, electricity 
is non-storable, which causes demand and supply to be balanced on a knife-edge. Relatively 
small changes in load or generation can cause large changes in price and all in a matter 
of hours, if not minutes. In this respect there is no other market like it. 

Californians are very well aware of this. In January 2001 California's energy market was 
on the verge of collapse. Wholesale electricity prices have soared since summer 2000,
see the top panel of Fig. 1. The state's largest utilities were threatening that they 
would be bankrupted unless they were allowed to raise consumer electricity rates by 30\%;
the California Power Exchange suspended trading and filed for Chapter 11 protection 
with the U.S. Bankruptcy Court.
How could this have happened when deregulation was supposed to increase efficiency and 
bring down electricity prices? It turns out that the difficulties that have appeared 
are intrinsic to the design of the market, in which demand exhibits virtually no price 
responsiveness and supply faces strict production constraints \cite{borenstein01,stoft01}. 

Another flaw of deregulation was the underestimation of the rising consumption of 
electricity in California. The soaring prices and San Francisco blackouts 
clearly showed that there is a need for sophisticated tools for the analysis of 
market structures and modeling of electricity load dynamics \cite{bll99,kaminski99}.
In this paper we investigate whether electricity loads in the California 
power market can be modeled by ARMA models.

\section{Preparation of the data}

The analyzed database was provided by the University of California Energy Institute (UCEI,
www.ucei.org). Among other data it contains system-wide loads supplied by California's 
Independent (Transmission) System Operator. This is a time series containing the load for 
every hour of the period April 1st, 1998 -- December 31st, 2000. Due to a very strong daily 
cycle we have created a 1006 days long sequence of daily loads. Apart from the daily cycle, 
the time series exhibits weekly and annual seasonality, see the bottom panel of Fig. 1. 
Because common trend and seasonality removal techniques do not work well when the time 
series is only a few (and not complete, in our case ca. 2.8 annual cycles) cycles long, 
we restricted the analysis only to two full years of data, i.e. to the period January 
1st, 1999 -- December 31st, 2000, and applied a new seasonality reduction technique 
\cite{wkn01}.

\begin{figure}[tbp]
\centerline{\epsfxsize=12cm \epsfbox{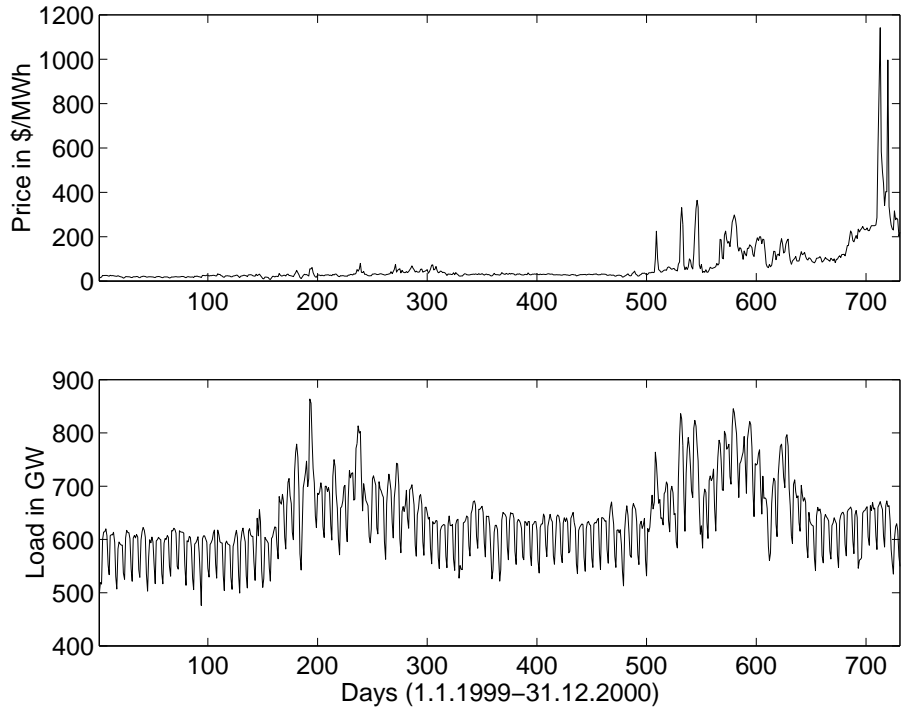}}
\caption{California Power Exchange daily average market clearing prices ({\it top panel}) 
and California power market daily system-wide load ({\it bottom panel}) since January 1st, 
1999 until December 31st, 2000. The annual and weekly seasonalities are clearly visible.}
\end{figure}

The seasonality can be easily observed in the frequency domain by plotting 
a sample analogue of the spectral density, i.e. the periodogram
\begin{equation}
I_n(\omega_k)=\frac{1}{n} \left|\sum_{t=1}^{n} x_t \exp\{-2\pi i
(t-1) \omega_k\} \right|^2,
\end{equation}
where $\{x_1,...,x_n\}$ is the vector of observations, $\omega_k = k/n$,
$k=1,...,[n/2]$ and $[x]$ denotes the largest integer less then or
equal to $x$. In the top panel of Fig. 2 we plotted the periodogram for
the system-wide load. It shows well-defined peaks at
frequencies corresponding to cycles with period 7 and 365 days.
The smaller peaks close to $\omega_k=0.3$ and 0.4 indicate periods
of 3.5 and 2.33 days, respectively. Both peaks are the so called
harmonics (multiples of the 7-day period frequency) and indicate
that the data exhibits a 7-day period but is not sinusoidal. The
weekly period was also observed in lagged autocorrelation plots
\cite{weron00}. 

\begin{figure}[tbp]
\centerline{\epsfxsize=12cm \epsfbox{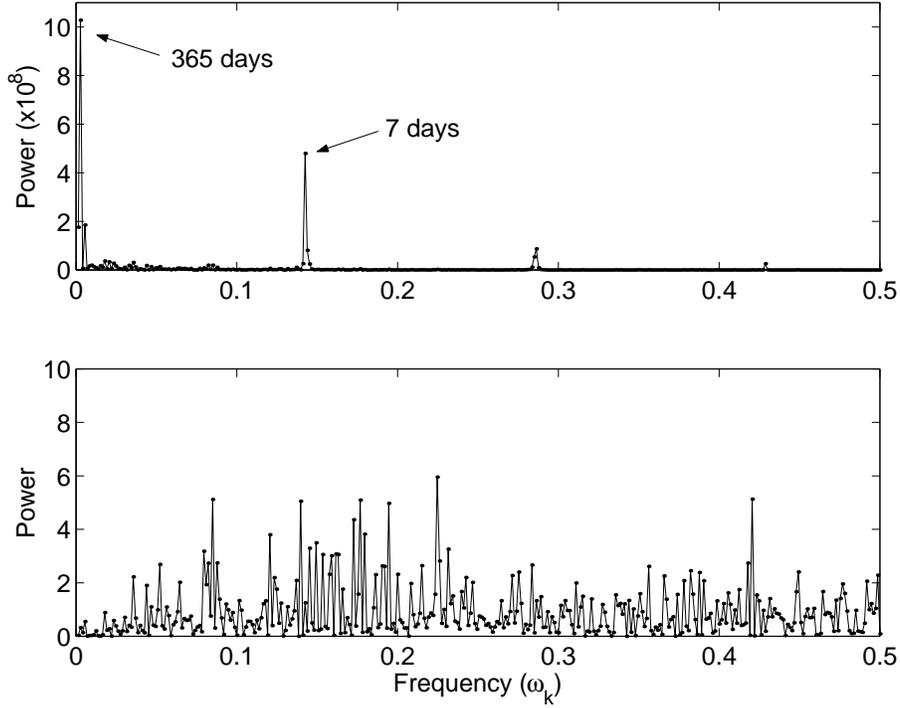}}
\caption{Periodogram of the California power market daily system-wide load since January 1st,
1999 until December 31st, 2000 ({\it top panel}). The annual and weekly frequencies are clearly
visible. Periodogram of the load returns after removal of the weekly and annual cycles 
({\it bottom panel}). No dominating frequency can be observed.}
\end{figure}

To remove the weekly cycle we used the moving average technique
\cite{bd96}. For the vector of daily loads $\{x_1,...,x_{731}\}$
the trend was first estimated by applying a moving average filter
specially chosen to eliminate the weekly component and to dampen
the noise:
\begin{equation}
\hat{m}_t=\frac17({x_{t-3}+...+x_{t+3}}),
\end{equation}
where $t=4, ..., 728$. Next, we estimated the seasonal component.
For each $k=1,...,7$, the average $w_k$ of the deviations
$\{(x_{k+7j}-\hat{m}_{k+7j}), 4\le k+7j\leq 728\}$ was computed.
Since these average deviations do not necessarily sum to zero, we
estimated the seasonal component $s_k$ as
\begin{equation}
\hat{s}_k = w_k-\frac17\sum_{i=1}^{7}w_i,
\end{equation}
where $k=1,...,7$ and $\hat{s}_k=\hat{s}_{k-7}$ for $k>7$. The
deseasonalized (with respect to the 7-day cycle) data was then
defined as
\begin{equation}
d_t=x_t-\hat{s}_t \quad \mbox{for $t=1,...,731$}.
\end{equation}
Finally we removed the trend from the deseasonalized data $\{d_t\}$ by taking logarithmic 
returns $r_t = \log (d_{t+1}/d_t)$, $t=1,...,730$. 

After removing the weekly seasonality we were left with the annual
cycle. Unfortunately, because of the short length of the time
series (only two years), the method applied to the 7-day cycle
could not be used to remove the annual seasonality. To overcome this we
applied a new method which consists of the following \cite{wkn01}:
\begin{description}
\item[(i)] calculate a 25-day rolling volatility \cite{kaminski97}
for the whole vector $\{r_1, ..., r_{730}\}$;
\item[(ii)] calculate the average volatility
for one year, i.e. in our case
\begin{equation}
v_t=\frac{v_t^{1999}+v_t^{2000}}{2};
\end{equation}
\item[(iii)] smooth the volatility by taking a 25-day moving average
of $v_t$;
\item[(iv)] finally, rescale the returns by dividing
them by the smoothed annual volatility.
\end{description}
The obtained time series (see the top panel of Fig. 3) showed
no apparent trend and seasonality (see the bottom panel of Fig. 2). 
Therefore we treated it as a realization of a stationary process. 
Moreover, the dependence structure exhibited only short-range correlations.
Both, the autocorrelation function (ACF) and the partial autocorrelation function 
(PACF) rapidly tend to zero (see the bottom panels of Fig. 3), which suggests that 
the deseasonalized load returns can be modeled by an ARMA-type process.

\section{Modeling with ARMA processes}

The mean-corrected (i.e. after removing the sample mean=0.0010658) deseasonalized load 
returns were modeled by ARMA (Autoregressive Moving Average) processes 
\begin{equation}
X_t-\phi_1 X_{t-1}-...-\phi_p X_{t-p} = Z_t+\theta_1
Z_{t-1}+...+\theta_q Z_{t-q}, \quad t=1,...,n,
\end{equation}
where $(p,q)$ denote the order of the model and $\{Z_t\}$ is a sequence of
independent, identically distributed variables with mean 0 and variance $\sigma^2$
(denoted by $\mbox{iid}(0,\sigma^2)$ in the text).

\begin{figure}[tbp]
\centerline{\epsfxsize=12cm \epsfbox{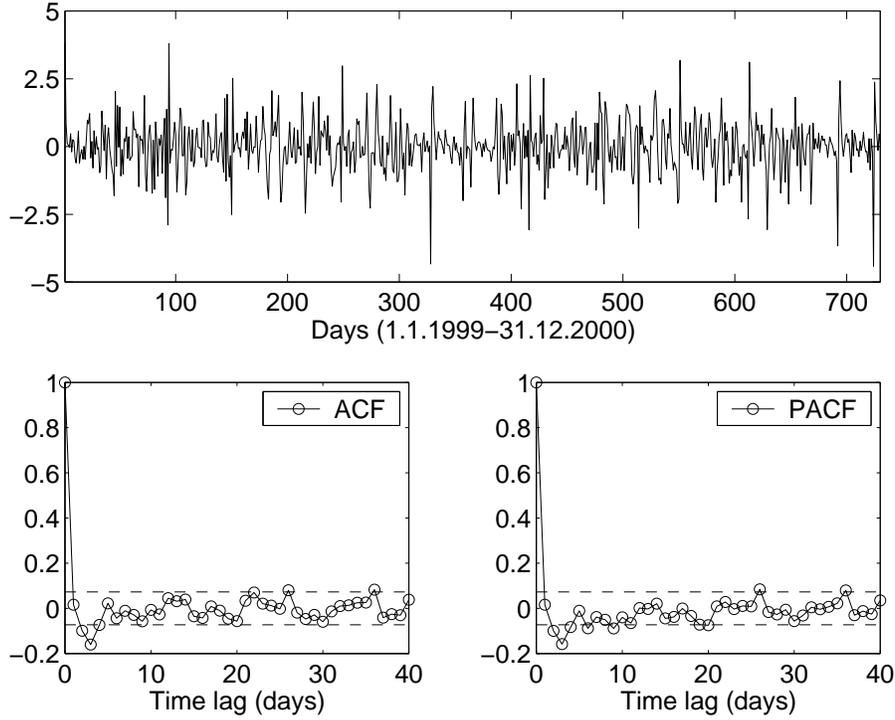}} 
\caption{Load returns after removal of the weekly and annual cycles ({\it top panel}).
The ACF ({\it bottom left panel}) and PACF ({\it bottom right panel}) for the mean-corrected 
deseasonalized load returns. Dashed lines represent the bounds $\pm 1.96/\sqrt{730}$, i.e. 
the 95\% confidence intervals of Gaussian white noise.}
\end{figure}

The maximum likelihood estimators $\hat{\phi}=(\hat{\phi_1},...,\hat{\phi_p})$, 
$\hat{\theta}=(\hat{\theta_1},...,\hat{\theta_q})$ and $\hat{\sigma}^2$ of the parameters
$\phi=(\phi_1,...,\phi_p)$, $\theta=(\theta_1,...,\theta_q)$ and $\sigma^2$, respectively, 
were obtained after a preliminary estimation via the Hannan-Rissanen method \cite{bd96}
using all 730 deseasonalized returns. 
The parameter estimates and the model size ($p$, $q$) were selected to be those that 
minimize the bias-corrected version of the Akaike criterion, i.e. the AICC statistics
\begin{equation}
  \mbox{AICC}=-2\ln L+\frac{2(p+q+1)n}{n-p-q-2},
\end{equation}
where $L$ denotes the maximum likelihood function and $n=730$.

The optimization procedure led us to the following ARMA(1,6) model (with $\theta_4=\theta_5=0$)
\begin{eqnarray}\label{ARMA}
  X_t 
& = & 0.332776X_{t-1} +\nonumber \\ 
&   & + Z_t-0.383245Z_{t-1}-0.12908Z_{t-2}-0.149307Z_{t-3}-0.0531862 Z_{t-6},
\end{eqnarray}
where $t=1,...,730$ and $\{Z_t\}\sim$ iid$(0,0.838716)$. The value of the AICC
criterion obtained for this model was AICC=1956.294.

In order to check the goodness of fit of the model to the set of data we compared 
the observed values with the corresponding predicted values obtained from the fitted model. 
If the fitted model was appropriate, then the residuals
\begin{equation}
\hat{W}_t = 
\frac{X_t-\hat{X}_t(\hat{\phi},\hat{\theta})}{\sqrt{\varsigma_{t-1}(\hat{\phi},\hat{\theta})}},
\quad t=1,...,730,
\end{equation}
where $\hat{X}_t(\hat{\phi},\hat{\theta})$ denotes the predicted
value of $X_t$ based on $X_1,...,X_{t-1}$ and $\varsigma_{t-1} = E(X_t - \hat X_t)^2/\sigma^2$, 
should behave in a manner that 
is consistent with the model. In our case this means that the properties of the residuals 
should reflect those of an iid noise sequence with mean 0 and variance $\sigma^2$. 

\begin{figure}[tbp]
\centerline{\epsfxsize=12cm \epsfbox{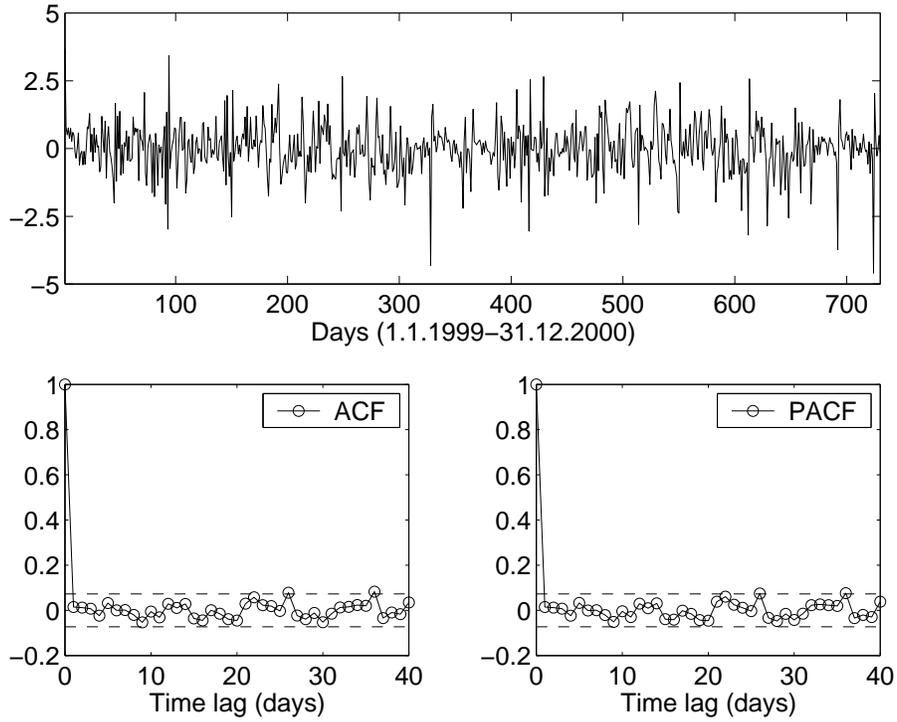}} 
\caption{The residuals obtained from the ARMA(1,6) model ({\it top panel}). 
The ACF ({\it bottom left panel}) and the PACF ({\it bottom right panel})
of the residuals. Dashed lines represent the bounds $\pm 1.96/\sqrt{730}$.}
\end{figure}

The residuals obtained from the ARMA(1,6) model fitted to the mean-corrected deseasonalized 
load returns are displayed in the top panel of Fig. 4. The graph gives no indication of a
nonzero mean or nonconstant variance.
The sample ACF and PACF of the residuals fall between the bounds $\pm 1.96/\sqrt{730}$ 
indicating that there is no correlation in the series, see the bottom panels of Fig. 4.
Recall that for large sample size $n$ the sample autocorrelations of an iid sequence 
with finite variance are approximately iid with distribution $N(0,1/n)$.
Therefore there is no reason to reject the fitted model on the basis of the autocorrelation 
or partial autocorrelation function. However, we should not rely only on simple visual 
inspection techniques. For our results to be more statistically sound we performed several 
standard tests for randomness. The results of the portmanteau, turning 
point, difference-sign and rank tests are presented in Table 1. 
Short descriptions of all applied tests can be found in the Appendix.

\begin{table}[tb]
\caption{Test statistics and p-values for the residuals.}
\begin{tabular}{lll}
  \hline
  Test ~~~~~~~~~~~~~~~~~~~~~ & Test statistics value & p-value \\
  \hline
  Portmanteau                & 15.03  & (0.7747) \\
  Turning point              & 464    & (0.0609) \\
  Difference-sign            & 361    & (0.6536) \\
  Rank                       & 131090 & (0.5529) \\
  \hline
\end{tabular}
\end{table}

As we can see from Table 1, if we carry out the tests at commonly used 5\% level, 
the tests do not detect any deviation from the iid behavior. Thus there is not
sufficient evidence to reject the iid hypothesis.  
Moreover, the order $p=0$ of the minimum AICC autoregressive model for 
the residuals also suggests the compatibility of the residuals with white noise,
see the Appendix.
Therefore we may conclude that the ARMA(1,6) model (defined by eq. (\ref{ARMA})) fits 
the mean-corrected deseasonalized load returns very well.

\section{Distribution of the residuals}

In the previous Section we showed that the residuals are a realization of an iid(0,$\sigma^2$)
sequence. But what precisely is their distribution? The answer to this question is 
important, because if the noise distribution is known then stronger conclusions can be drawn
when a model is fitted to the data. There are simple visual inspection techniques that enable
us to check whether it is reasonable to assume that observations from an iid sequence are 
also Gaussian. The most widely used is the so-called normal probability plot, see Fig. 5. 
If the residuals were Gaussian
then they would form a straight line. Obviously they are not Gaussian -- the deviation 
from the line is apparent. This deviation suggests that the residuals have heavier tails.

\begin{figure}[tbp]
\centerline{\epsfxsize=12cm \epsfbox{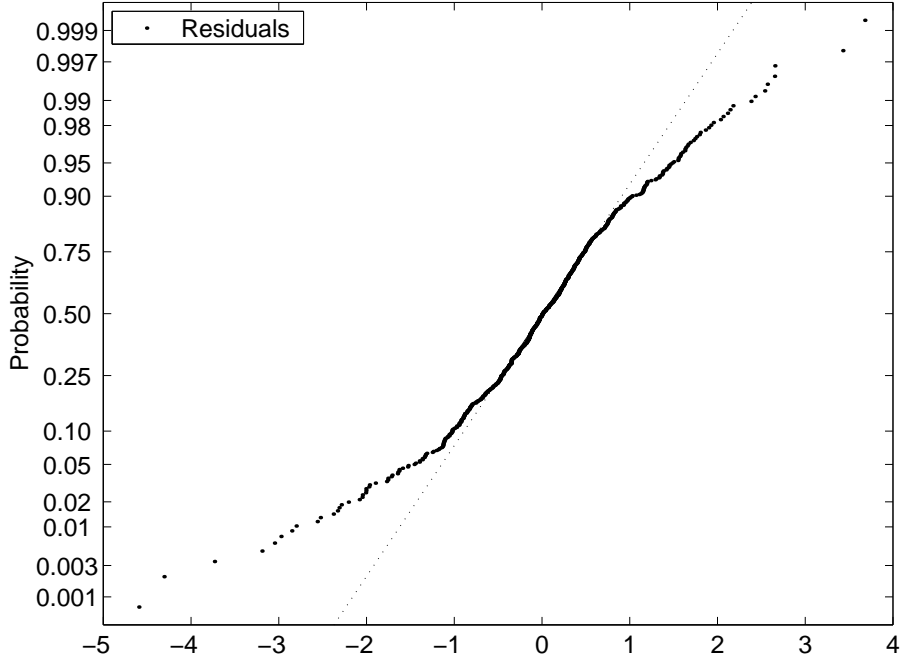}} 
\caption{The normal probability plot of the residuals obtained from the ARMA(1,6) model.
If the residuals were Gaussian then they would form a straight line.}
\end{figure}

However, we have to bear in mind that in order to comply with the ARMA model 
assumptions the distribution of the residuals must have a finite second moment. 
In the class of heavy-tailed laws 
with finite variance the hyperbolic distribution seems to be a natural candidate.

The hyperbolic law was introduced by Barndorff-Nielsen \cite{b-n77}
for modeling the grain size distribution of windblown sand. It was also found to provide 
an excellent fit to the distributions of daily returns
of stocks from a number of leading German enterprises \cite{ek95,knss94}. 
The name of the distribution is derived from the fact that its log-density forms a hyperbola. 
Recall that the log-density of the normal distribution is a parabola. Hence the hyperbolic 
distribution provides the possibility of modeling heavy tails.

The hyperbolic distribution is defined as a normal variance-mean mixture where 
the mixing distribution is the Inverse Gaussian law. More precisely, a random 
variable has the hyperbolic distribution if its density is of the form 
\begin{equation}\label{denhyp} 
f(x; \alpha,\beta,\delta,\mu) 
  = \frac{\sqrt{\alpha^2 - \beta^2}}{2\alpha\delta K_1(\delta\sqrt{\alpha^2 - \beta^2})}
    \exp\left\{ -\alpha \sqrt{\delta^2+(x-\mu)^2} + \beta(x-\mu) \right\},
\end{equation} 
where the normalizing constant 
$\mbox{K}_{1}(t)=\frac12 \int_0^{\infty} \exp\{-\frac12 t (x+\frac1x)\} dx$, $t>0$, 
is the modified Bessel function with index 1, the scale parameter $\delta>0$, 
the location parameter $\mu \in R$ and $0\le |\beta|<\alpha$. The latter two parameters --
$\alpha$ and $\beta$ -- determine the shape, with $\alpha$ being responsible for 
the steepness and $\beta$ for the skewness.

Given a sample of independent observations all four parameters can be estimated by 
the maximum likelihood method. In our studies we used the 'hyp' program \cite{bs92}
to obtain the following estimates 
$$
\hat\alpha = 1.671304, \qquad \hat\beta = -0.098790, \qquad \hat\delta = 0.298285, 
\qquad \hat\mu = 0.076975.
$$
The empirical probability density function (PDF) -- to be more precise: a kernel estimator 
of the density -- together with the estimated hyperbolic PDF are presented in Fig. 6. 
We can clearly see that, on the semi-logarithmic scale, the tails of the residuals' 
density form straight lines, which justifies our choice of the theoretical distribution. 
The adjusted Kolmogorov statistics $K=\sqrt{n} \sup_x |F(x)-F_n(x)|$, where $F(x)$ is the 
theoretical and $F_n(x)$ is the empirical cummulative distribution function, returns 
the value $K=1.5652$. This indicates that there is not
sufficient evidence to reject the hypothesis of the hyberbolic distribution of the 
residuals at the 1\% level. For comparison we fitted a Gaussian law to the residuals as well.
In this case the adjusted Kolmogorov statistics returned $K=1.8019$ causing us to reject 
the Gaussian hypothesis of the residuals at the same level.

\begin{figure}[tbp]
\centerline{\epsfxsize=12cm \epsfbox{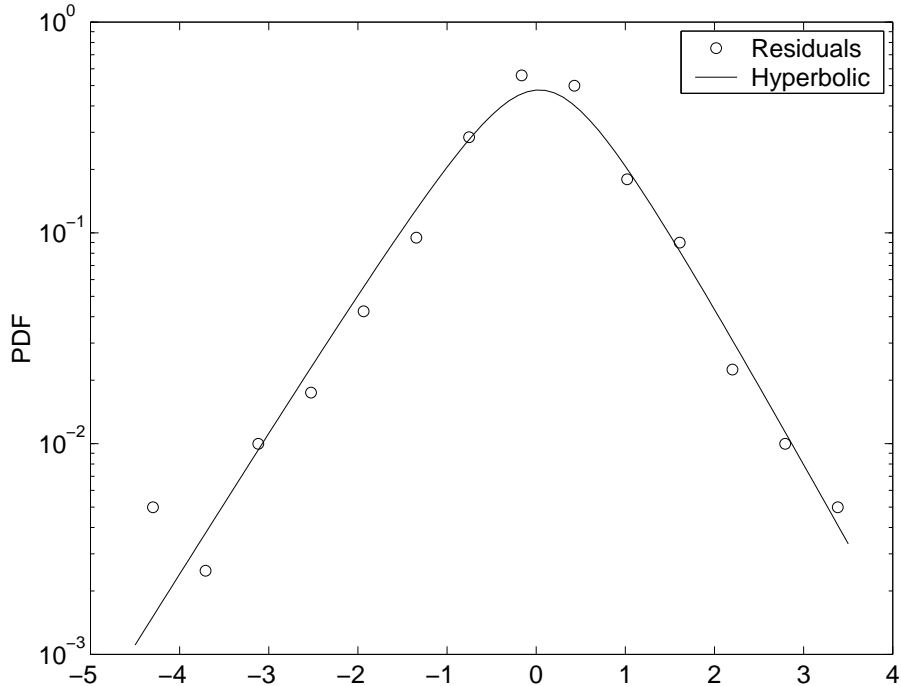}} 
\caption{The empirical probability density function (a kernel estimator of the density) 
and the approximating hyperbolic PDF on the semi-logarithmic scale. 
}
\end{figure}

\section{Conclusions}

Due to limited monitoring in a power distribution system its loads usually are not known
in advance and can only be forecasted based on the available information. In this paper
we showed that it is possible to model deseasonalized loads via ARMA processes
with heavy-tailed hyperbolic noise. This method could be used to forecast loads
in a power market. Its effectiveness, however, still has to be tested and will be
the subject of our further research.

\section*{Appendix: Tests for randomness}

\begin{description}

\item[The portmanteau test.] Instead of checking to see if each sample autocorrelation 
  $\hat{\rho}(j)$ falls inside the bounds $\pm 1.96/\sqrt{n}$, where $n$ is the sample size, 
  it is possible to consider a single statistic introduced by Ljung and Box \cite{lb78}
  $Q=n(n+2)\sum_{j=1}^{h}\hat{\rho}^2(j)/(n-j)$,
  whose distribution can be approximated by the $\chi^2$
  distribution with $h$ degrees of freedom.
  A large value of $Q$ suggests that the sample autocorrelations of the observations are
  too large for the data to be a sample from an iid sequence.
  Therefore we reject the iid hypothesis at level $\alpha$ if
  $Q>\chi^2_{1-\alpha}(h)$, where $\chi^2_{1-\alpha}$ is the $(1-\alpha)$ quantile of
  the $\chi^2$ distribution with $h$ degrees of freedom.

\item[The turning point test.] If $y_1,...,y_n$ is a sequence of
  observations, we say that there is a turning point at time $i$
  ($1<i<n$) if $y_{i-1}<y_i$ and $y_i>y_{i+1}$ or if $y_{i-1}>y_i$ and
  $y_i<y_{i+1}$. In order to carry out a test of the iid
  hypothesis (for large $n$) we denote the number of turning
  points by $T$ ($T$ is approximately $N(\mu_T,\sigma^2_T)$, where
  $\mu_T=2(n-2)/3$ and $\sigma^2_T=(16n-29)/90$) and we reject
  this hypothesis at level $\alpha$ if
  $|T-\mu_T|/\sigma_T>\Phi_{1-\alpha/2}$, where
  $\Phi_{1-\alpha/2}$ is the $(1-\alpha/2)$ quantile of the
  standard normal distribution. The large value of $T-\mu_T$
  indicates that the series is fluctuating more rapidly than
  expected for an iid sequence; a value of $T-\mu_T$ much smaller
  than zero indicates a positive correlation between neighboring
  observations.

\item[The difference-sign test.] For this test we count the
  number $S$ of values $i$ such that $y_i>y_{i-1}$, $i=2,...,n$.
  For an iid sequence and for large $n$, $S$ is approximately
  $N(\mu_S,\sigma^2_S)$, where $\mu_S=(n-1)/2$ and
  $\sigma^2_S=(n+1)/12$. A large positive (or negative) value of
  $S-\mu_S$ indicates the presence of an increasing (or
  decreasing) trend in the data. We therefore reject the
  assumption of no trend in the data if
  $|S-\mu_S|/\sigma_S>\Phi_{1-\alpha/2}$.

\item[The rank test.] The rank test is particularly useful for
  detecting a linear trend in the data. We define $P$ as the
  number of pairs $(i,j)$ such that $y_j>y_i$ and $j>i$,
  $i=1,...,n-1$. For an iid sequence and for large $n$, $P$ is
  approximately $N(\mu_P,\sigma^2_P)$, where $\mu_P=n(n-1)/4$ and
  $\sigma^2_P=n(n-1)(2n+5)/72$. A large positive (negative) value
  of $P-\mu_P$ indicates the presence of an increasing
  (decreasing) trend in data. The iid hypothesis is therefore
  rejected at level $\alpha$ if $|P-\mu_P|/\sigma_P>\Phi_{1-\alpha/2}$.

\item[The minimum AICC AR model test.]
  A simple test for whiteness of a time series is to fit autoregressive models of
  orders $p=0,1,...,p_{max}$, for some large $p_{max}$, and to record the value of 
  $p$ for which the AICC value attains the minimum. Compatibility of these observations 
  with white noise is indicated by selection of the value $p=0$. 

\end{description}

\end{document}